\documentclass[conference]{IEEEtran}
\IEEEoverridecommandlockouts

\usepackage{tikz}
\usepackage{pgfplots}
\pgfplotsset{compat=newest}
\usepgfplotslibrary{groupplots,dateplot}
\usetikzlibrary{patterns,shapes.arrows}

\usepackage{algorithm}
\usepackage[noend]{algpseudocode}
\usepackage{xspace}
\usepackage{multirow}
\usepackage{tabularx}
\usepackage{makecell}
\usepackage{adjustbox}
\usepackage{float}

\usepackage{mathtools}
\usepackage{listings}
\usepackage{comment}
\usepackage{ragged2e}
\usepackage{framed}
\usepackage{upquote}
\usepackage{soul}

\def\BibTeX{{\rm B\kern-.05em{\sc i\kern-.025em b}\kern-.08em
    T\kern-.1667em\lower.7ex\hbox{E}\kern-.125emX}}
\begin{document}

\title{Terastal: Layer-Variant-based Scheduling for Real-Time
Multi-DNN Workloads on Heterogeneous Accelerators
\thanks{This is the author accepted manuscript of a paper accepted by the 32nd IEEE International Conference on Embedded and Real-Time Computing Systems and Applications (RTCSA 2026). The definitive version will appear in the IEEE proceedings. \copyright~2026 IEEE. Personal use of this material is permitted. Permission from IEEE must be obtained for all other uses, in any current or future media, including reprinting/republishing this material for advertising or promotional purposes, creating new collective works, for resale or redistribution to servers or lists, or reuse of any copyrighted component of this work in other works.}
}

\author{ \IEEEauthorblockN{Sing-Yao Wu, Fengshuo Song, Eli Bozorgzadeh} \IEEEauthorblockA{ Department of Computer Science, University of California, Irvine, CA, USA} }
\maketitle

\begin{abstract}
Heterogeneous DNN accelerators improve soft real-time multi-DNN execution by mapping each layer to its preferred accelerator to reduce latency. However, under skewed workloads, large layer-latency differences across accelerators limit scheduling flexibility and increase deadline misses.
To address this challenge, we introduce \emph{layer variants}, customized layer implementations that reduce latency gaps on non-preferred accelerators. We then present \emph{Terastal}, a soft real-time framework for layer-variant design and scheduling on heterogeneous DNN accelerators. \emph{Terastal} combines offline heterogeneity-aware virtual budget assignment and layer-variant design, and online scheduling to jointly optimize accelerator mapping and variant selection under timing and accuracy constraints. Experimental results show that \emph{Terastal} reduces deadline miss rate per model by $40.58\%$, $30.53\%$, and $36.27\%$ compared with FCFS, EDF, and DREAM~\cite{ref:kim23dream}, respectively, while incurring only $2.24\%$ average normalized accuracy loss across models with variants.
\end{abstract}

\begin{IEEEkeywords}
Heterogeneous DNN accelerators, soft real-time scheduling, multi-DNN workloads, layer variants
\end{IEEEkeywords}

\section{Introduction}
Modern edge systems increasingly rely on hardware accelerators to meet real-time requirements of computation-intensive deep neural network (DNN) workloads.
To meet timing constraints, recent designs have moved from programmable accelerators, such as GPUs and FPGAs, to DNN-customized ASIC accelerators. To support diverse edge DNNs, recent platforms increasingly integrate accelerators customized for different layers.
However, this heterogeneity introduces new scheduling challenges for real-time multi-DNN workloads.

Multi-DNN workloads exhibit highly heterogeneous execution latency because different layers have different affinities to accelerator architectures and dataflows. For example, layers with many channels are more efficient on weight-stationary (WS) dataflow accelerators~\cite{ref:nvdla17}, while layers with large output feature maps are more efficient on output-stationary (OS) dataflow accelerators~\cite{ref:du15shidiannao}. Executing a layer on a \textit{non-preferred} accelerator can substantially increase latency and reduce the chance of meeting model deadlines.
To better match diverse layers to suitable hardware, recent architectures integrate multiple heterogeneous accelerators into a single platform~\cite{ref:kwon21herald,ref:zeng22h3m,ref:odema24scar}. However, designing a heterogeneity-aware scheduler for real-time multi-DNN workloads remains challenging. Recent work, such as DREAM~\cite{ref:kim23dream}, brings heterogeneity awareness to layer-wise real-time DNN scheduling. Even with such awareness, under \textit{skewed workloads}---where many ready layers simultaneously target the same accelerator type---some layers must either wait
or execute on non-preferred accelerators, increasing deadline misses.

A natural mitigation is to assign each layer a time budget or virtual deadline~\cite{ref:hong11locldeadline, ref:babaei24sgprs, ref:babaei25daris}, allowing slack-rich layers to run on non-preferred accelerators with low deadline violation risk. However, layer-wise budgeting alone is often insufficient. Some layers remain too slow on non-preferred accelerators to meet the assigned budget, limiting remapping benefits.
For these layers, offline layer fine-tuning can sometimes reduce the latency on non-preferred accelerators, producing optimized implementations, which we call \textit{layer variants}.
At runtime, these variants can be applied when dispatched to non-preferred accelerators to reduce latency to meet budgets. Yet, layer variants introduce two challenges. First, each variant may incur per-layer accuracy loss, and using multiple variants in one inference can accumulate unacceptable degradation. Second, generating variants for all layer–accelerator pairs incurs prohibitive storage overhead. Therefore, layer variants must be selected carefully and coordinated with runtime scheduling.

To address this need, we present \emph{Terastal}, a soft real-time framework that jointly designs layer variants offline and schedules them online to reduce deadline misses for multi-DNN workloads on heterogeneous accelerators. Offline, \emph{Terastal} decomposes each model deadline into layer-wise virtual budgets, identifies latency-critical layers whose execution on non-preferred accelerators would hinder deadline satisfaction, and selectively designs variants only for those layers
to limit storage overhead. At runtime, \emph{Terastal} uses budget-derived virtual deadlines and layer-wise slack to jointly decide accelerator mapping and variant usage, reducing deadline misses while controlling accuracy degradation. 
Experiments show that \emph{Terastal} reduces average per-model deadline miss rate by $40.58\%$, $30.53\%$, and $36.27\%$ over FCFS~(First-Come-First-Served), EDF (Earliest-Deadline-First), and DREAM~\cite{ref:kim23dream}, respectively, with only $2.24\%$ average normalized accuracy loss on models with layer variants.

Our main contributions are summarized as follows:
1) We introduce \textit{layer variants}, customized layer implementations that reduce latency on non-preferred accelerators.
2) We propose a budget-guided framework for layer variant construction and virtual-budget/virtual-deadline assignment.
3) We design an online scheduler that uses layer-wise slack to jointly choose accelerator mappings and layer variants, reducing deadline misses while respecting accuracy constraints.

\section{Related Work}
Prior work on DNN accelerator scheduling addresses different aspects of the problem. Several studies~\cite{ref:kao22magma, ref:odema24scar, ref:odema25hydra-tesla} rely on offline scheduling and thus adapt poorly to workload variations at runtime. Other works~\cite{ref:kwon21herald,ref:zhou25taichi} develop online schedulers, but do not explicitly account for deadline constraints.
Real-time approaches~\cite{ref:choi20prema, ref:ghodrati20planaria, ref:kim23moca, ref:blanco24relmas} address timing requirements, but often either ignore accelerator heterogeneity or rely on heavier optimization mechanisms that are less suitable for resource-constrained edge systems. 
DREAM~\cite{ref:kim23dream} considers accelerator heterogeneity in real-time scheduling, but provides limited layer-wise timing insight, making it difficult to support layer variant integration.
BlastNet~\cite{ref:ling22blastnet} and DARIS~\cite{ref:babaei25daris} optimize scheduling for CPU--GPU or GPU-only systems, and therefore their platform-specific block design and virtual deadline mechanisms do not directly generalize to heterogeneous DNN accelerator platforms.
Approaches based on anytime/early-exit inference~\cite{ref:rahmath2024early, ref:yao20scheduling, ref:soyyigit22anytime} are related but orthogonal to our work.
These approaches reduce inference cost by adapting model execution depth or components, whereas our work focuses on runtime scheduling and layer mapping on heterogeneous accelerators under real-time constraints. To the best of our knowledge, {\em Terastal} is among the first frameworks to integrate selective layer-level variants with soft real-time scheduling for heterogeneous multi-DNN accelerator platforms.

\section{Layer Variant Architecture}
A layer variant reduces the latency of executing a layer on a non-preferred accelerator by reshaping the computation to better match the target dataflow. On heterogeneous platforms, DNN accelerators primarily use output-stationary (OS)~\cite{ref:du15shidiannao} or weight-stationary (WS)~\cite{ref:nvdla17} dataflows. OS dataflow keeps partial sums local and typically parallelizes over output activations, making it efficient for layers with large output feature maps. WS dataflow keeps filter weights local and parallelizes over weights, making it more efficient for layers with many channels. Because different DNN layers vary in spatial and channel size, the same layer can run significantly slower on a non-preferred accelerator~\cite{ref:kwon21herald}.

To mitigate this, we design layer variants using the space-to-depth (S2D) and depth-to-space (D2S) transformations~\cite{ref:yang19deeperlab}. These operations redistribute information across spatial and channel dimensions while preserving the layer’s input-output functionality at the model level. Intuitively, S2D folds spatial structure into channels, reducing feature-map resolution while increasing channel count, whereas D2S unfolds channels back into spatial dimensions. Together, these transformations preserve tensor-shape compatibility while reformulating the intermediate computation for a target accelerator dataflow.

\begin{figure}[t]
    \centering
    \includegraphics[width=0.85\linewidth]{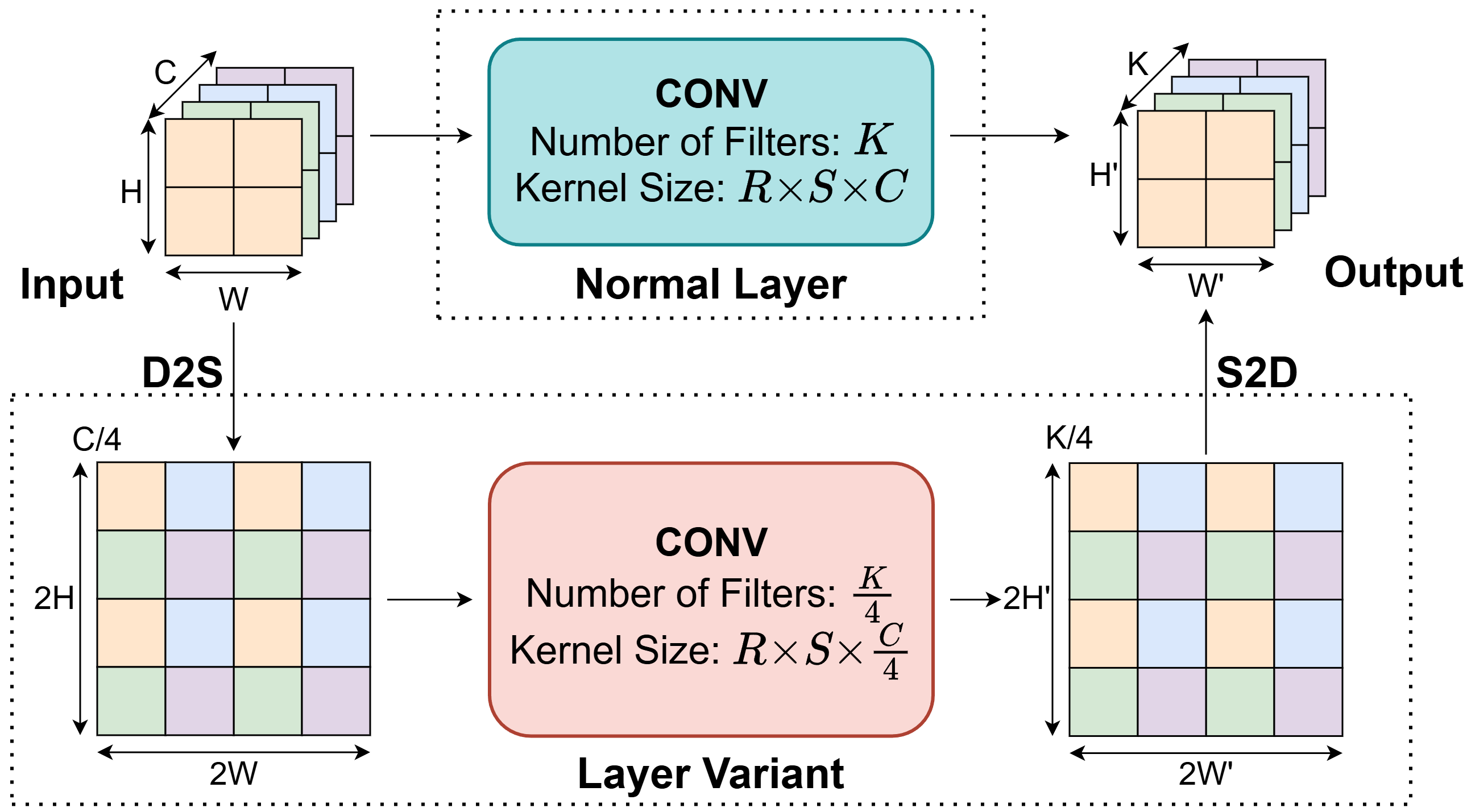}
    \vspace{-2ex}
    \caption{Example of converting a WS-preferred convolution into a layer variant better suited for OS execution when $\gamma=2$.}
    \label{fig:layer-variant-design}
    \vspace{-2ex}
\end{figure}
Figure~\ref{fig:layer-variant-design} illustrates this idea using a convolution layer with $K$ filters of size $(R\times S \times C)$, and input feature map of size $(H\times W \times C)$, where $H$ and $W$ are the input spatial dimensions, $R$ and $S$ are the filter spatial dimensions, and $C$ is the number of channels. Suppose the layer is better suited for WS execution, and we aim to reduce its latency on an OS accelerator. We first apply a D2S transformation with ratio $\gamma > 1$ to unfold channels into the spatial dimensions, changing the input tensor shape to $(\gamma H \times \gamma W \times \frac{C}{\gamma^{2}})$, assuming $C$ is divisible by $\gamma^{2}$. The variant convolution then operates on this transformed tensor using $\frac{K}{\gamma^{2}}$ filters, each of size $R \times S \times \frac{C}{\gamma^{2}}$. Compared with the original layer, each filter in the variant operates on only $\frac{C}{\gamma^2}$ input channels instead of $C$, while the output feature map becomes $\gamma^2$ times larger in the spatial dimensions. This increases output-side parallelism and PE utilization, making the variant more suitable for OS accelerators. Moreover, because both the number of input channels per filter and the number of filters are reduced by a factor of $\gamma^2$, the total number of weights in the variant is reduced by $\frac{1}{\gamma^4}$, helping limit storage overhead. After convolution, we apply an S2D operation with the same $\gamma$ to fold the spatial dimensions back into the channel dimension and restore the original output shape.
This principle also extends to other layer types when they can be represented in convolution-equivalent form. For instance, a fully connected layer is a case of convolution in which the kernel covers the full input spatial dimensions.

However, a layer variant is only an approximation of the original layer, and overusing variants can degrade accuracy. Therefore, variants should be designed and applied selectively and used only when the timing benefit justifies the associated accuracy and storage cost. Based on this observation, we next present the Terastal framework, which uses virtual budgets to determine which layers should receive variants and when the online scheduler applies them at runtime.

\section{Terastal Framework}
\begin{figure}
    \centering
    \vspace{-1ex}
    \includegraphics[width=0.9\linewidth]{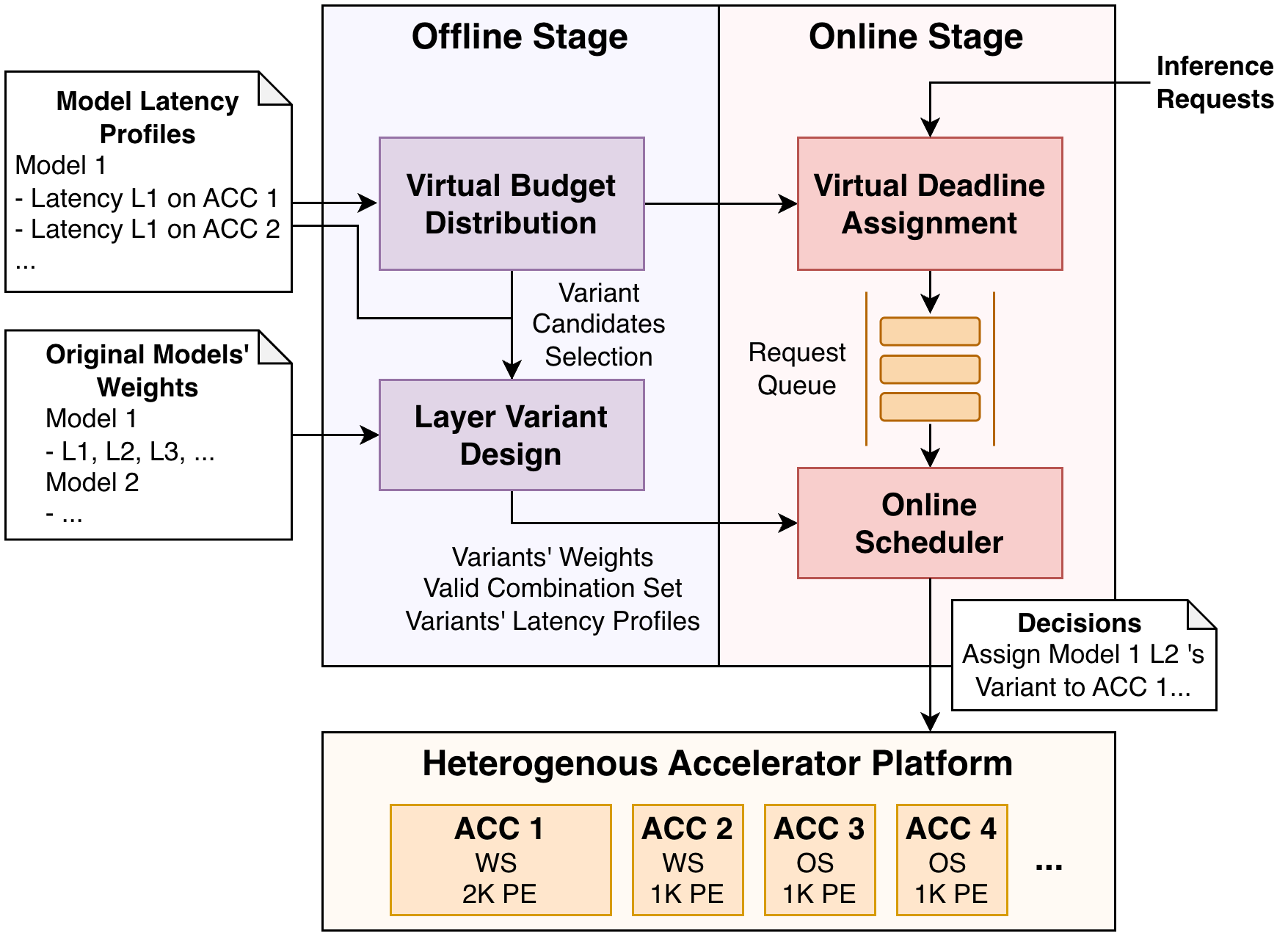}
    \vspace{-2ex}
    \caption{Terastal Framework Overview}
    \label{fig:system-overview}
    \vspace{-3ex}
\end{figure}

Figure~\ref{fig:system-overview} shows the overview of our system. We consider an edge platform with \(n_a\) heterogeneous accelerators that differ in PE size and dataflow, denoted by the set
\(\mathcal{A}=\{A_1,A_2,\ldots,A_{n_a}\}\).
All accelerators share an on-chip memory, allowing consecutive layers of the same inference to execute on different accelerators. We assume layer-granularity scheduling, where each layer is modeled as a non-preemptive job. This is consistent with prior DNN accelerator scheduling work~\cite{ref:kim23dream, ref:kwon21herald} and avoids additional hardware and software support required for intra-layer preemption~\cite{ref:choi20prema}. Scheduling decisions are therefore made only at layer boundaries. After a layer on \(A_k\) finishes, its output activation is written back to the shared memory.
The system serves a fixed set of \(n_m\) DNN models
\(\mathcal{M}=\{M_1,\dots,M_{n_m}\}\).
Each model \(M_m\) consists of \(L_m\) layers, indexed by
\(\ell\in\{1,\dots,L_m\}\). Each layer $\ell$ takes its previous layer’s output as input. If layer $\ell$ has a variant, we denote this variant by $\hat{\ell}$.
We denote the execution latency of layer $\ell$ in model $M_m$ on accelerator $A_k$ by $c_{m,\ell,k}$. At runtime, the $j$-th request of model $M_m$ is denoted as $J_{j,m}$. $J_{j,m}$ arrives at time $t^a_{j,m}$ and has a relative deadline $D_m$. A request misses its deadline if it does not finish by $t^a_{j,m} + D_m$. Since late completion has limited benefit, our objective is to reduce deadline misses under soft real-time multi-DNN workloads.

\label{sec:framework}

\emph{Terastal} consists of an offline stage and an online stage. Offline, the virtual budget distribution module decomposes each model deadline into layer-wise virtual budgets. Guided by these budgets, the layer variant design module identifies latency-critical layers and selectively constructs variants that reduce latency on non-preferred accelerators while controlling accuracy loss. Online, the scheduler uses the offline-generated budgets, layer latency profiles, and valid variant combinations to assign a virtual deadline to each ready layer and jointly decide accelerator mapping and variant usage. Thus, \emph{Terastal} tightly couples offline variant generation with online deadline-aware scheduling. We next describe each module in detail.

\subsection{Virtual Budget Distribution} \label{sec:virtual-budget-distribution}
We first assign \textit{layer-wise virtual budgets} to decompose end-to-end deadlines into layer-level timing guidance. These budgets are used both offline, to decide where variants are worth designing, and online, to compute per-layer urgency.

For a model $M_m$ with relative deadline $D_m$, we assign a \emph{virtual budget}, $b_{m, \ell}$, to each layer $\ell$ in $M_m$, such that:
\begin{equation}
\label{eq:sum-vd-eq-D}
    \sum_{\ell=1}^{L_m} b_{m, \ell} = D_m
\end{equation}

The \emph{virtual deadline} of layer $\ell$ of request $J_{j,m}$ during runtime is computed as:
\begin{equation}
\label{eq:def-vd}
   d^v_{j,m,\ell} = t^a_{j,m} + \sum_{\ell'=1}^{\ell} b_{m, \ell'}
\end{equation}

Thus, each layer receives an intermediate soft deadline, enabling the scheduler to reason about per-layer urgency when deciding execution time and accelerator mapping.

Because DNN accelerators are highly deterministic, per-layer latency on a given accelerator is usually fixed and can be profiled offline.
For each layer $\ell$ of each $M_m$, we collect its execution latencies $\{\,c_{m,\ell,k}\,\}_{k\in\mathcal{A}}$ and sort the \emph{distinct} values in strictly decreasing order, giving:
$\,c_{m,\ell}^{\downarrow(1)} > c_{m,\ell}^{\downarrow(2)} > \cdots > c_{m,\ell}^{\downarrow(R_{m,\ell})}\,$,
where $c_{m,\ell}^{\downarrow(r)}$ is the $r$-th largest execution time and $R_{m,\ell}$ is the number of unique latencies.

A natural way to assign virtual budgets is to distribute $D_m$ proportionally to each layer’s worst-case latency $c_{m, \ell}^{\downarrow(1)}$, i.e.,

\begin{equation}
    b_{m, \ell} = D_m \times \frac{c_{m, \ell}^{\downarrow(1)}}{\sum_{\ell'=1}^{L_m} c_{m, \ell'}^{\downarrow(1)}}
\end{equation}

However, this is often infeasible on heterogeneous platforms. Because cross-accelerator latency differences can be large, in many cases, $\sum_{\ell'=1}^{L_m} c_{m, \ell'}^{\downarrow(1)}>D_m $, so proportional assignment yields budgets smaller than each layer’s worst-case latency, and sometimes even its minimum achievable latency, making the resulting virtual deadlines unattainable. Prior work assigns budgets based on maximum recent latency~\cite{ref:babaei25daris}, but since it targets GPUs, it does not address this issue.

Our key insight is that a layer budget need not accommodate all accelerators. If $b_{m, \ell}$ lies in the interval $c^{\downarrow(r)}_{m, \ell}>b_{m, \ell} \ge c^{\downarrow(r+1)}_{m, \ell}$, then accelerators with latency $c_{m, \ell}^{\downarrow(r)}$ are excluded, while those with latency at most $c_{m, \ell}^{\downarrow(r+1)}$ remain feasible.
At runtime, a slower accelerator is used only when earlier layers have accumulated enough slack for the current layer to still meet its virtual deadline. Otherwise, layer variants can be applied to reduce the latency gap.

\begin{algorithm}[!t]
\caption{Offline Layer-Wise Virtual Budget Distribution}
\label{alg:vd-assign}
\begin{algorithmic}[1]
\Statex \textbf{Input:} Model $M_m$, Deadline $D_m$
\Statex \textbf{Output:} Layer-wise budget set $\{b_{m,1}, b_{m,2}, ..., b_{m,L_m}\}$
  \State Initialize $\rho_{m, \ell} \gets 1$ for all $\ell \in \{1,\dots,L_m\}$
    \While{$true$}
        \State $C_{total} \gets \sum_{\ell=1}^{L_{m}} c_{m, \ell}^{\downarrow(\rho_{m, \ell})}$
        \If{$C_{total} \le D_m $} \Comment{Budget validity check}
            \For{each layer $\ell$ \textbf{in} $M_m$} \label{ln:vb_assign-assign-start}
                \State $b_{m, \ell} \gets D_m \times \frac{c_{m, \ell}^{\downarrow (\rho_{m, \ell})}}{C_{total}}$
            \EndFor \label{ln:vb_assign-assign-end}
            \State \Return $\{b_{m,1}, b_{m,2}, ..., b_{m,L_m}\}$
        \EndIf
        \State $\mathcal{S} \gets \{\, \ell \in M_m \mid \rho_{m, \ell} < R_{m,\ell} \,\}$ \label{ln:vb_assign-tight-start}
        \If{$\mathcal{S} = \emptyset$} \Comment{Feasibility check}
          \State \Return \textbf{Fail}
        \EndIf
        \State $\ell^* \gets \operatorname*{arg\,max}_{\ell \in \mathcal{S}}
                \bigl(c_{m,\ell}^{\downarrow(\rho_{m, \ell})} - c_{m,\ell}^{\downarrow(\rho_{m, \ell}+1)}\bigr)$
        \State $\rho_{m, \ell^*} \gets \rho_{m, \ell^*} + 1$ \label{ln:vb_assign-tight-end}

    \EndWhile
\end{algorithmic}
\end{algorithm}

Based on this idea, Algorithm~\ref{alg:vd-assign} assigns layer-wise virtual budgets through a constraint level $\rho_{m,\ell} \in \{1,\dots,R_{m,\ell}\}$ for each layer. At level $\rho_{m,\ell}$, all accelerators with latency larger than $c^{\downarrow(\rho_{m,\ell})}_{m,\ell}$ are \emph{excluded} from the budget assignment. The algorithm first attempts to assign budgets proportionally using the currently selected latency level $c^{\downarrow(\rho_{m, \ell})}_{m, \ell}$ of each layer~(lines~\ref{ln:vb_assign-assign-start}--\ref{ln:vb_assign-assign-end}).
If the resulting total latency exceeds $D_m$, it greedily tightens one layer at a time by selecting the layer with the largest gap to its next lower latency level and increasing its constraint level by $1$. This process continues until the total assigned latency fits within $D_m$~(lines~\ref{ln:vb_assign-tight-start}--\ref{ln:vb_assign-tight-end}). If all layers already use their fastest latency level and the total still exceeds $D_m$, model $M_m$ is infeasible on platform $\mathcal{A}$.

The resulting constraint levels and virtual budgets directly guide layer-variant design. Layers with high constraint levels, especially those with a large latency gap between adjacent levels, are strong candidates for variant construction because their feasible accelerator choices are highly restricted.
For each selected layer, we design a variant that targets accelerators whose original latency is at least $c^{\downarrow(r)}_{m, \ell}$ and reduces their latency to at most $c^{\downarrow(r+1)}_{m, \ell}$.
In this way, variants expand the set of accelerators on which a layer can meet its assigned budget.

\subsection{Layer Variant Design}
\label{sec:lv-design}
Section~\ref{sec:virtual-budget-distribution} identifies layers that are too constrained for non-preferred execution; thus, we design variants for these latency-critical layers. We profile per-layer latency across accelerators with offline cost-analysis tools~\cite{ref:kwon20maestro} and select layers whose non-preferred execution latencies exceed their budgets. For each candidate layer, we choose the minimum $\gamma$ that reduces latency on the target non-preferred accelerator to the next constraint level, or below the latency of the preferred accelerator. Each variant is trained independently by replacing the original layer and freezing all other layers. We also profile the latency $c_{m, \hat{\ell}, k}$ of every variant $\hat{\ell}$ of layer $\ell$ of $M_m$ on each accelerator $A_k$, and provide these to the online scheduler.

\begin{figure}
    \centering
    \includegraphics[width=\linewidth]{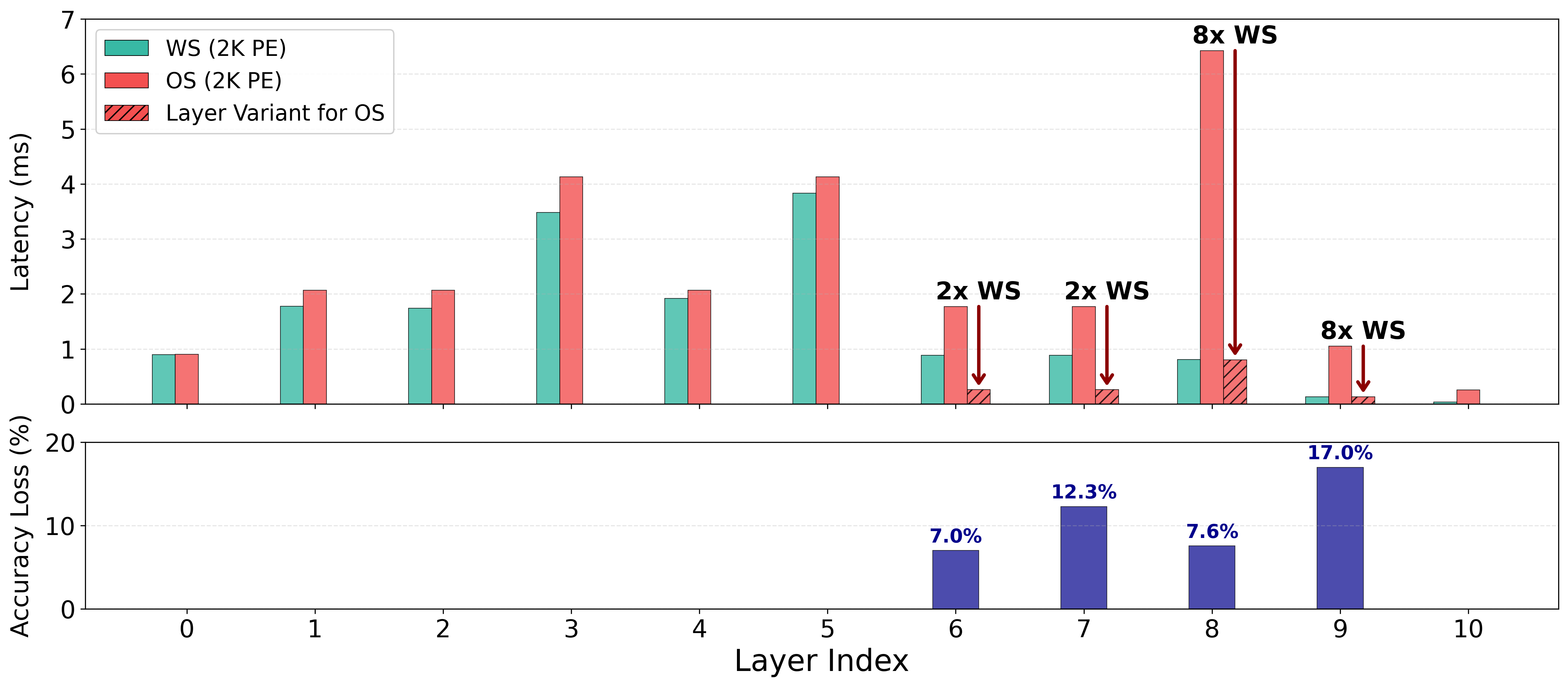}
    \vspace{-5ex}
    \caption{Per-layer latency of VGG11 on WS and OS accelerators (top) and accuracy loss per individual layer variant (bottom).}
    \label{fig:vgg-layerwise-latency}
    \vspace{-4ex}
\end{figure}

Figure~\ref{fig:vgg-layerwise-latency} illustrates this process on VGG11. The top plot compares per-layer latency on WS and OS accelerators with the same number of PEs. While the early layers have similar WS/OS latency, the later layers incur $2\times$ to $8\times$ higher latency on OS, indicating a strong affinity for WS. These layers are therefore selected as candidates for OS-oriented variants. After variant design, their OS latency becomes comparable to or lower than that on WS, showing the latency reduction on non-preferred accelerators. The bottom plot shows that individual variants incur $7.0\%$–$17.0\%$ accuracy loss due to fewer weights than the original layers. This loss is layer-dependent and compounds when multiple variants are applied in one inference.

To control accuracy loss, we assign each model an accuracy threshold $\theta_m$. We evaluate all feasible variant combinations offline and pass the valid set $\mathcal{V}_m$ to the scheduler, which admits only those whose offline-measured accuracy exceeds $\theta_m$.

\subsection{Online Scheduler Design}
Using the virtual budgets from Section~\ref{sec:virtual-budget-distribution} and the variant profiles and valid combination set from Section~\ref{sec:lv-design}, the online scheduler jointly decides accelerator mapping and variant usage at runtime. For each request, the scheduler first computes the virtual deadline of every ready layer using Eq.~\ref{eq:def-vd}. For a ready layer $\ell$ of request $J_{j,m}$ (i.e., all its preceding layers have completed), let $\tau_k(t)=t+w_k(t)$ denote the next available time of accelerator $A_k$, where $w_k(t)$ is the remaining execution time of the layer currently running on $A_k$ at time $t$. The estimated finish time on accelerator $A_k$ at time $t$ is:
\begin{equation}
\label{eqn:finish-time-def}
    t^f_{j,m,\ell \to k}(t) = \tau_k(t) + c_{m,\ell,k},
\end{equation}

If layer $\ell$ has a variant $\hat{\ell}$, its estimated finish time on accelerator $A_k$ is:
\begin{equation}
\label{eqn:varian-finish-time-def}
    t^f_{j,m,\hat{\ell} \to k}(t) = \tau_k(t) + c_{m,\hat{\ell},k},
\end{equation}

During each scheduling round, after assigning a layer to $A_k$, we update $\tau_k(t)$ to that layer’s estimated finish time so that subsequent decisions account for both the currently executing layer and earlier assignments in the same round.

The slack of a layer at time $t$ when assigning to $A_k$ is defined as:
\begin{equation}
\label{eqn:slack-def}
    s_{j,m,\ell\to k}(t) = d^v_{j,m, \ell} -t^f_{j,m, \ell\to k}(t)
\end{equation}

Because latency variation across accelerators can be large, worst-case slack is too pessimistic to reflect urgency. We therefore measure a layer’s urgency by its best-case slack across accelerators, defined as:
\begin{equation}
\label{eqn:optimal-slack-def}
    s^*_{j,m,\ell}(t) = \max_k {s_{j,m,\ell\to k}(t)}
\end{equation}
A smaller best-case slack indicates a more urgent layer.

\begin{algorithm}[!t]
\caption{Online Scheduling Algorithm}
\label{alg:online-sched}
\begin{algorithmic}[1]
\State $\mathcal{A}_{idle}\gets\{\text{idle accelerators in }\mathcal{A}\}$
\State $\mathcal{J}_{ready}\gets\{\text{all ready request--layer pairs } (J_{j,m}, \ell)\}$

\For{$(J_{j,m},\ell)$ \textbf{in} $\mathcal{J}_{ready}$ \textbf{sorted by} $s^*_{j, m,\ell}(t)$ \textbf{ascending}}\label{ln:online-sched-stage1-start} 
\Statex \Comment{Assign to an accelerator that meets its virtual deadline}
\State $\mathcal{A}_{cands}\gets\{\,A_k\in\mathcal{A}_{idle}\mid t^{f}_{j,m,\ell\to k}(t)\le d^{v}_{j,m,\ell}\,\}$ \label{ln:online-sched-vd_normal_cand-start} 
\If{$\mathcal{A}_{cands}\neq\emptyset$}
  \State $A_k^* \gets \arg\min_{A_k\in\mathcal{A}_{cands}}\, t^{f}_{j,m,\ell\to k}(t)$
  \State \textbf{assign } $\ell$ \textbf{ of } $J_{j,m}$ \textbf{ to } $A_{k}^*$
\State $\mathcal{A}_{idle} \gets \mathcal{A}_{idle} \setminus \{A_k^*\}$
\State $\mathcal{J}_{ready}\gets\mathcal{J}_{ready}\setminus \{(J_{j,m}, \ell)\}$
\State \textbf{continue}
\EndIf \label{ln:online-sched-vd_normal_cand-end} 

\Statex \Comment{Apply a variant that can meet its virtual deadline}

\If{\textsc{LayerVariantFeasible}$(J_{j,m}, \ell)$}\label{ln:online-sched-vd_lv_cand-start} 
  \State $\mathcal{A}_{cands}\gets\{\,A_k\in\mathcal{A}_{idle} \mid t^{f}_{j,m,\hat{\ell}\to k}(t)\le d^{v}_{j,m,\ell}\,\}$

  \If{$\mathcal{A}_{cands}\neq\emptyset$}
    \State $A_k^* \gets \arg\min_{A_k\in\mathcal{A}_{cands}}\, t^{f}_{j,m,\hat{\ell}\to k}(t)$
    \State \textbf{assign } $\hat{\ell}$ \textbf{ of } $J_{j,m}$ \textbf{ to } $A^*_{k}$
    \State $\mathcal{A}_{idle} \gets \mathcal{A}_{idle} \setminus \{A_{k}^*\}$
    \State $\mathcal{J}_{ready}\gets\mathcal{J}_{ready}\setminus \{(J_{j,m}, \ell)\}$
    \State \textbf{continue}
  \EndIf
\EndIf \label{ln:online-sched-vd_lv_cand-end}
\EndFor \label{ln:online-sched-stage1-end}
  \Statex \Comment{Assign remaining idle accelerators based on slack gain}
\If{$\mathcal{A}_{idle}\neq\emptyset$ \textbf{ and }$\mathcal{J}_{ready}\neq\emptyset$}\label{ln:online-sched-stage2-start}
\For{$A_k$ \textbf{in} $\mathcal{A}_{idle}$}
\State $(J_{j,m}^*, \ell^*) = \arg\max_{(J_{j,m}, \ell)\in\mathcal{J}_{ready}}\, \Delta s_{j, m,\ell\to k}(t)$
\State \textbf{assign } $\ell^*$ \textbf{ of } $J_{j,m}^*$ \textbf{ to } $A_{k}$
\State $\mathcal{J}_{ready}\gets\mathcal{J}_{ready}\setminus \{(J_{j,m}^*,\ell^*)\}$
\EndFor
\EndIf \label{ln:online-sched-stage2-end}
\end{algorithmic}
\end{algorithm}

Algorithm~\ref{alg:online-sched} describes our online scheduling algorithm, which consists of two stages. In the first stage, the scheduler sorts all ready layers by non-decreasing best-case slack and prioritizes the most urgent layer that can still meet its virtual deadline on some idle accelerator. If one or more idle accelerators can complete the original layer before its virtual deadline, the scheduler assigns the layer to the accelerator with the earliest finish time (lines~\ref{ln:online-sched-vd_normal_cand-start}--\ref{ln:online-sched-vd_normal_cand-end}). Otherwise, if the layer has a variant and applying it keeps the current request within the valid combination set $\mathcal{V}_m$, the scheduler evaluates the variant finish times and assigns the variant to the earliest-finishing feasible accelerator~(lines~\ref{ln:online-sched-vd_lv_cand-start}--\ref{ln:online-sched-vd_lv_cand-end}). This stage ensures that the most urgent request-layer pair is served first while respecting both timing and accuracy constraints.

If idle accelerators remain after the first stage, the second stage~(lines~\ref{ln:online-sched-stage2-start}--\ref{ln:online-sched-stage2-end}) backfills them to improve utilization. For mapping layer $\ell$ of $J_{j,m}$ to $A_k$, we define the future potential slack as: 
\begin{equation}
\label{eqn:future-slack-estimation}
    s^{f*}_{j,m,\ell\to k}(t) = d^v_{j,m, \ell+1} -t^f_{j,m, \ell\to k}(t)-\min_{k'}{c_{m, \ell+1, k'}}
\end{equation}
and the corresponding slack increase as:
\begin{equation}
\label{eqn:slack-gain}
\Delta s_{j,m,\ell\to k}(t) = s^{f*}_{j,m,\ell\to k}(t) - s^*_{j,m,\ell}(t)    
\end{equation}
This quantity estimates how much additional slack the request is expected to gain for its next layer if the current layer is executed on $A_k$. In the second stage, each remaining idle accelerator $A_k$ is assigned the layer with the highest $\Delta s_{j,m,\ell\to k}(t)$. If both the original layer and its variant are feasible under the accuracy threshold, the scheduler chooses the one with the larger slack increase.

The scheduler is invoked whenever an accelerator becomes idle. With 
$n_J$ ready layers and $n_a$ accelerators, the first stage costs $O(n_J \cdot n_a + n_J \log n_J)$, while the second stage costs
$O(n_J \cdot n_a^2)$. The total per-invocation complexity is therefore $O(n_J \cdot n_a^2 + n_J \log n_J)$. Since edge platforms typically use only a small number of accelerators, this overhead remains lightweight relative to layer execution time. We also apply an early-drop policy: if the current time plus the minimum possible latency of all remaining layers already exceeds a request’s absolute deadline, the request is dropped to avoid wasted work and free resources for other requests.

\section{Evaluation}
\subsection{Evaluation Settings}
\newcolumntype{L}{>{\raggedright\arraybackslash}X}

\begin{table}[h]
\centering
\vspace{-2ex}
\caption{Accelerator Hardware Settings in Our Evaluation}
\vspace{-2ex}
\label{tab:hardware-settings}
\renewcommand{\arraystretch}{1.2}
\setlength{\tabcolsep}{4pt}
\begin{tabularx}{\linewidth}{|c|L|L|}
\hline
\textbf{\shortstack{\#PEs}} & \textbf{Dataflow / PE partition} & \textbf{Scenario set} \\ \hline

\multirow{1}{*}{4K}
  & \makecell[l]{1 WS (2K) + 2 OS (1K each)\\1 OS (2K) + 2 WS (1K each)}
  & \makecell[l]{AR social interaction\\AR gaming (light)\\Multi-Camera Vision (light)} \\ \hline

\multirow{1}{*}{6K}
  & \makecell[l]{1 WS (2K) + 2 OS (2K each)\\1 OS (2K) + 2 WS (2K each)}
  & \makecell[l]{AR social interaction\\AR gaming (heavy)\\Multi-Camera Vision (heavy)} \\ \hline

\end{tabularx}
\vspace{-3ex}
\end{table}

\begin{table}[h]
\centering
\caption{Workload Scenarios in Our Evaluation. Models marked with * have layer variants.}
\vspace{-2ex}
\label{tab:workload-settings}
\renewcommand{\arraystretch}{1.15}
\begin{tabular}{|c|l|}
\hline
\textbf{Scenario} & \textbf{Used Models (\textit{FPS})} \\
\hline

\multirow{2}{*}{\shortstack{AR Social\\Interaction}}
  & FBNet-C~\cite{ref:wu19fbnet} (\textit{60}), Hand S/P~\cite{ref:ge19handsp} (\textit{30, Prob: $0.5$}),\\
  & Sp2Dense\textsuperscript{*}\cite{ref:ma18sparsetodense} (\textit{30}), MobileNetV2-SSD\textsuperscript{*}\cite{sandler18mobilenetv2} (\textit{30}) \\ \hline
\multirow{2}{*}{\shortstack{AR Gaming\\(Light)}}
  & Hand S/P~\cite{ref:ge19handsp} (\textit{30}), 
  PlaneRCNN~\cite{ref:liu19planercnn} (\textit{10}),\\
  & Sp2Dense\textsuperscript{*}\cite{ref:ma18sparsetodense} (\textit{30}), MobileNetV2-SSD\textsuperscript{*}\cite{sandler18mobilenetv2} (\textit{30}) \\ \hline
\multirow{2}{*}{\shortstack{AR Gaming\\(Heavy)}}
  & Hand S/P~\cite{ref:ge19handsp} (\textit{45}), 
  PlaneRCNN~\cite{ref:liu19planercnn} (\textit{15}),\\
  & Sp2Dense\textsuperscript{*}\cite{ref:ma18sparsetodense} (\textit{30}), MobileNetV2-SSD\textsuperscript{*}\cite{sandler18mobilenetv2} (\textit{45})\\ \hline
\multirow{3}{*}{\shortstack{Multi-Camera\\Vision\\(Light)}}
  & MobileNetV2-SSD\textsuperscript{*}\cite{sandler18mobilenetv2} (\textit{45}),\\
  & ResNet50\textsuperscript{*}\cite{ref:he16resnet} (\textit{15}), VGG11\textsuperscript{*}\cite{ref:simonyan14vgg} (\textit{15}),\\
  & InceptionV3\textsuperscript{*}\cite{ref:szegedy16inception} (\textit{15}), 
  Swin-Tiny\textsuperscript{*}\cite{ref:liu21swin} (\textit{10}) \\ \hline
\multirow{3}{*}{\shortstack{Multi-Camera\\Vision\\(Heavy)}}
  & MobileNetV2-SSD\textsuperscript{*}\cite{sandler18mobilenetv2} (\textit{60}),\\
  & ResNet50\textsuperscript{*}\cite{ref:he16resnet} (\textit{30}), VGG11\textsuperscript{*}\cite{ref:simonyan14vgg} (\textit{30}),\\
  & InceptionV3\textsuperscript{*}\cite{ref:szegedy16inception} (\textit{15}), 
  Swin-Tiny\textsuperscript{*}\cite{ref:liu21swin} (\textit{30}) \\ \hline
\end{tabular}
\vspace{-2ex}
\end{table}

We evaluate {\em Terastal} using a simulator built on MAESTRO~\cite{ref:kwon20maestro} and XRBench~\cite{ref:kwon23xrbench}. Table~\ref{tab:hardware-settings} summarizes the hardware configurations, and Table~\ref{tab:workload-settings} summarizes the workload scenarios. We consider heterogeneous platforms with 4K or 6K total PEs, where each PE is a MAC unit, and vary the partitioning across WS~\cite{ref:nvdla17} and OS~\cite{ref:du15shidiannao} accelerators. All accelerators use $8$ MiB shared on-chip memory, $128$ GB/s off-chip bandwidth, and $1$~GHz clock frequency.
We match each workload scenario to a hardware setting with an appropriate total PE count, avoiding trivial all-pass or all-fail cases. 
Each model is invoked as a periodic task with a period and relative deadline both equal to $(1/\text{FPS})$ seconds. The workloads include AR scenarios adapted from XRBench~\cite{ref:kwon23xrbench}, which contain a mix of models with and without layer variants, and Multi-Camera Vision scenarios representing a shared edge platform serving different concurrent vision applications from multiple end devices, in which all models include layer variants.

For models with layer variants, we select candidate layers using the profiling and virtual-budget method in Section~\ref{sec:framework}. For each selected layer, $\gamma$ is set to the minimum integer that reduces latency on the non-preferred accelerator to at or below that of the preferred accelerator. Variants are trained on the same datasets as their original models, including ImageNet~\cite{ref:russakovsky15imagenet}, Pascal VOC~\cite{ref:everingham15pascal}, and KITTI~\cite{ref:geiger13kitti}. The accuracy threshold for each model with layer variants is set to $90\%$ of its baseline accuracy without variants. The resulting storage overhead is modest, increasing per-model storage by only $0.5\%$ to $5.9\%$ relative to the original model sizes.

We compare our scheduler with FCFS, EDF, and DREAM~\cite{ref:kim23dream}.
FCFS prioritizes ready layers by arrival time, while EDF prioritizes them by their derived deadlines based on minimum execution time. Both map each selected layer to the idle accelerator with the lowest execution latency.
For DREAM, we replace its original objective---deadline miss rate multiplied by normalized energy---with deadline miss rate alone for fair comparison with \emph{Terastal}. We further include two ablation variants: Terastal--no budgeting, which applies layer variants but computes slack using EDF-style deadlines without virtual budgets, and Terastal--no variants, which uses virtual budgets for scheduling but disables layer variants. The same early request-drop policy is applied to all schedulers.

\subsection{Experimental Results}
\subsubsection{Latency and Accuracy Impact of Layer Variants}
\begin{figure}
    \centering
    \includegraphics[width=0.95\linewidth]{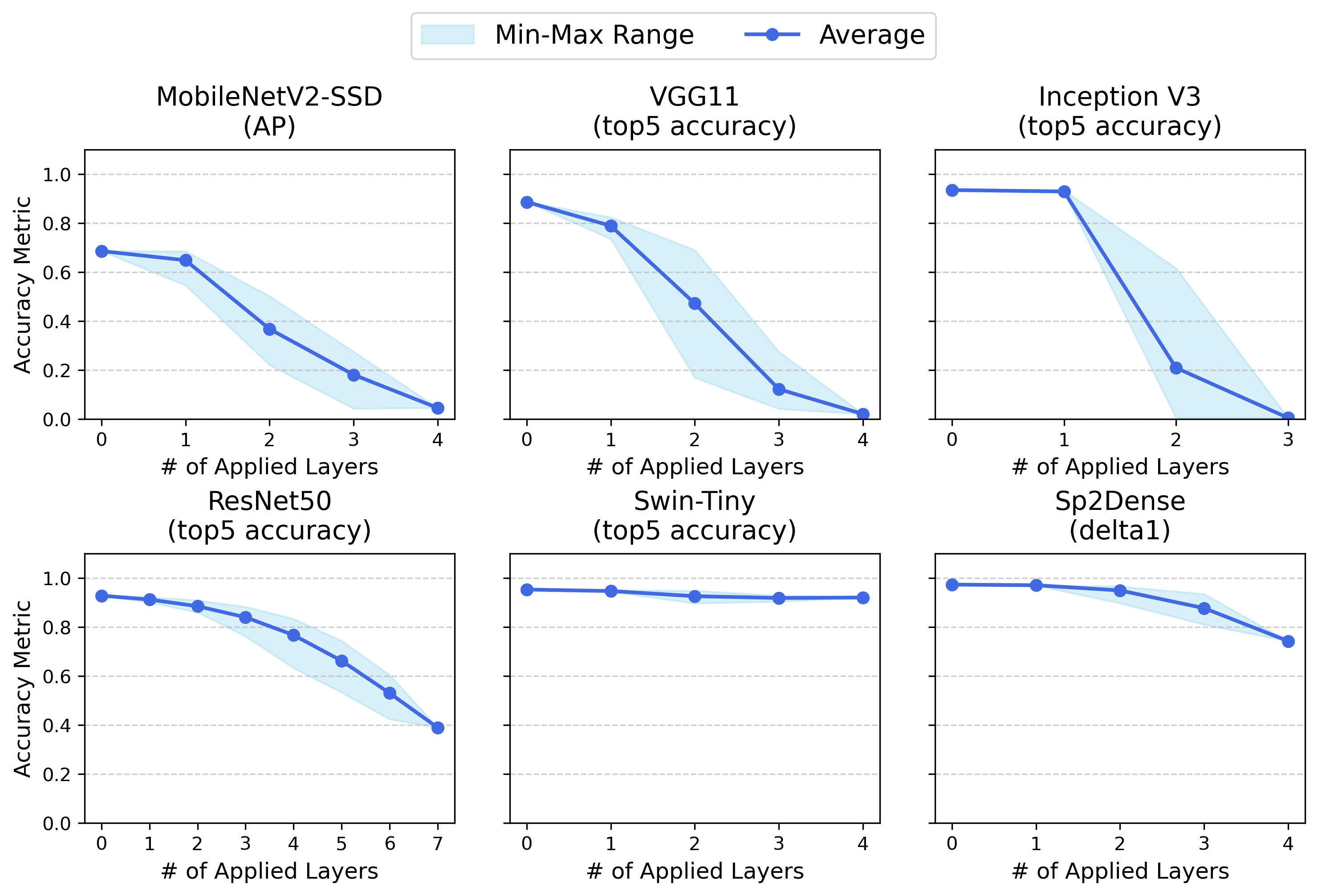}
    \vspace{-2ex}
    \caption{Accuracy metrics (AP, top-5 accuracy, etc.) versus the number of applied layer variants per inference, showing the mean and min--max range across all layer combinations with the same number of applied variants.}
    \label{fig:lv-accuracy-trend}
    \vspace{-2ex}
\end{figure}

We first examine whether layer variants provide useful timing flexibility. Across all hardware settings, layer variants with $\gamma \in \{2, 3\}$ reduce latency on non-preferred accelerators to at or below that of the preferred accelerators, allowing variant-enabled layers to meet their virtual budgets on more accelerator choices when variants are applicable. This benefit also extends to accelerators with the same dataflow but different PE sizes, showing that a variant designed for one accelerator can generalize across hardware instances of the same type.

Figure~\ref{fig:lv-accuracy-trend} shows the model accuracy under different combinations of applied layer variants. Since accuracy depends not only on the number of variants applied but also on which specific layers are modified, we report the mean and min-max range over all layer combinations with the same variant count. The range width reflects per-layer sensitivity: wider ranges indicate that certain layer positions or combinations are substantially more accuracy-sensitive. Models with higher architectural redundancy, such as ResNet50, Swin-Tiny, and Sp2Dense, remain robust under multiple variants, while models with more compact architectures are more sensitive even when only one or two variants are applied.

\subsubsection{Deadline Miss Rate Comparison}
\begin{figure}
    \centering
\includegraphics[width=1\linewidth]{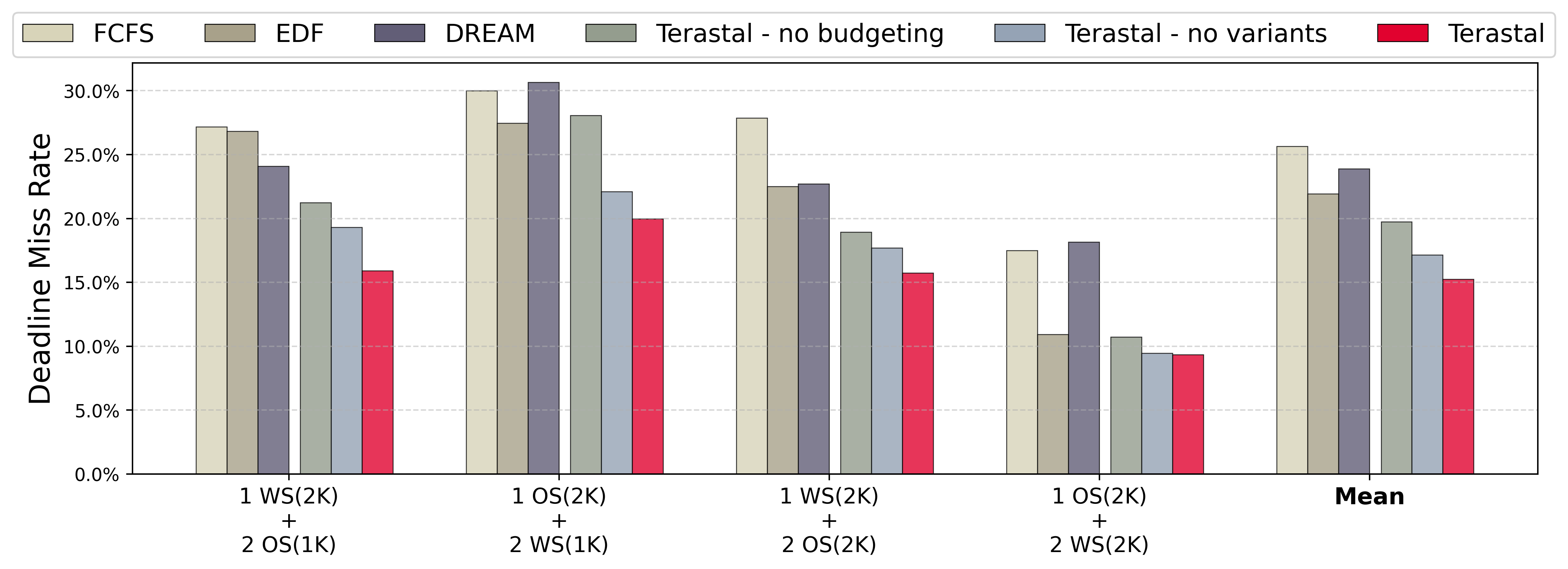} 
    \vspace{-5ex}
    \caption{Average per-model deadline miss rate comparison in different hardware settings}
    \label{fig:cross-test-set-dmr}
    \vspace{-3ex}
\end{figure}

Figure~\ref{fig:cross-test-set-dmr} shows the average of per-model deadline miss rates under each hardware setting. {\em Terastal} achieves the lowest overall mean miss rate, reducing average per-model deadline miss rate by $40.58\%$, $30.53\%$, and $36.27\%$ relative to FCFS, EDF, and DREAM, respectively. These gains are achieved while incurring only $2.24\%$ average normalized accuracy loss (normalized by baseline accuracy of each model).
The ablation study shows that virtual budgets and layer variants play complementary roles. Terastal--no variants consistently outperforms the baseline schedulers, indicating that virtual budgets alone improve urgency estimation and scheduling effectiveness. Full Terastal further improves over Terastal--no variants, showing that layer variants provide additional benefit on top of virtual budgeting. This benefit is smaller in the 1 OS (2K) + 2 WS (2K) setting, where two WS accelerators already alleviate queueing on the preferred accelerators since most layers are more efficient on WS in our workloads.
In contrast, Terastal--no budgeting is less effective: although it still outperforms the conventional baselines in several settings, it remains worse than the two budgeted versions, suggesting that layer variants alone cannot fully realize their benefit without accurate virtual budgets.

\subsubsection{Analysis of Accuracy Threshold Impact}
\begin{figure}
    \centering
    \includegraphics[width=0.9\linewidth]{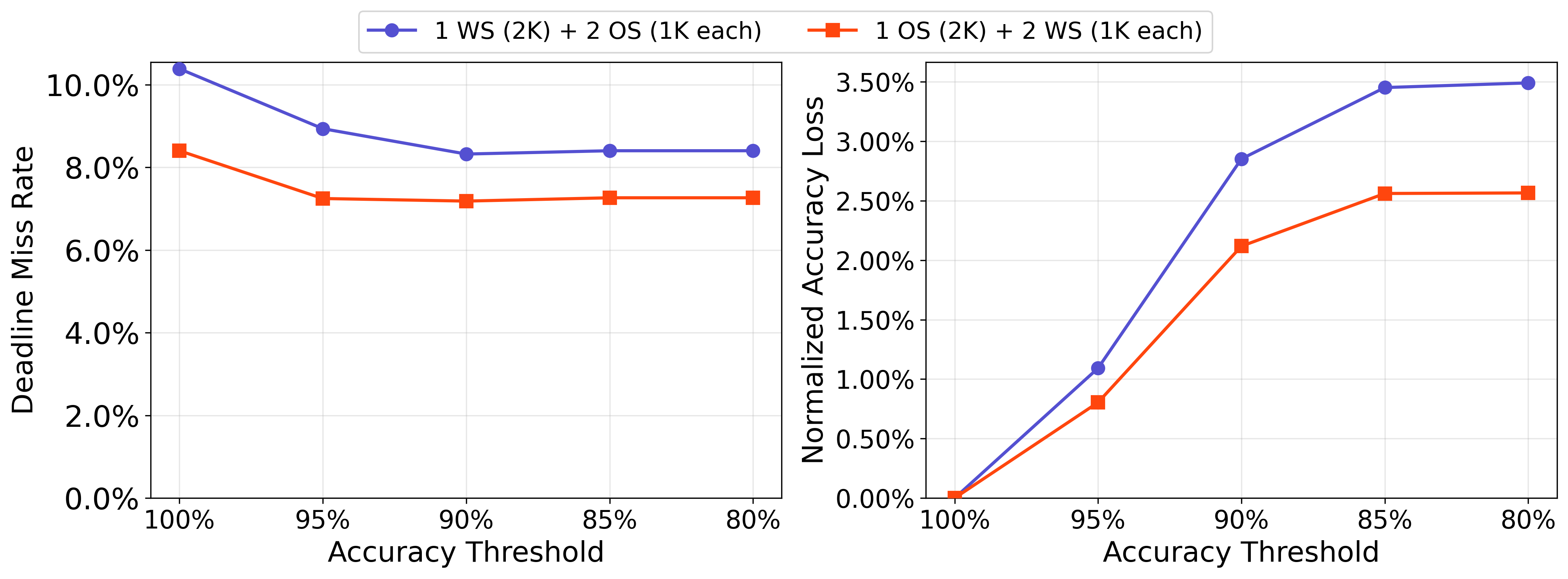}
    \vspace{-2ex}
    \caption{Deadline miss rate and normalized accuracy loss under different accuracy thresholds in Multi-Camera Vision (Light) workloads}
    \label{fig:quality-threshold-comparison}
    \vspace{-3ex}
\end{figure}

Figure~\ref{fig:quality-threshold-comparison} shows the impact of accuracy thresholds under two hardware settings for the Multi-Camera Vision (Light) workload. A threshold of $80\%$ allows model accuracy to drop to $80\%$ of baseline, whereas $100\%$ disallows any layer variants. Since most layers in this workload are more efficient on WS accelerators, the configuration with only one WS accelerator has a higher average per-model deadline miss rate when variants are not allowed. As the threshold is lowered, the miss-rate gap between the two hardware settings narrows, indicating that layer variants help balance the skewed workload by using non-preferred accelerators efficiently. Meanwhile, normalized accuracy loss increases, but remains within the threshold.

\section{Conclusion}
\label{sec:conclusion}
Heterogeneous DNN accelerators benefit real-time multi-DNN workloads by mapping layers to preferred accelerators to minimize latency. However, under skewed workloads, large cross-accelerator latency gaps make deadline-aware scheduling difficult. We presented \emph{Terastal}, a framework that combines virtual budget distribution,  layer-variant design, and online scheduling to selectively run customized pretrained layers on non-preferred accelerators and reduce deadline misses. Experimental results show that \emph{Terastal} reduces deadline miss rate by $40.58\%$, $30.53\%$, and $36.27\%$ relative to FCFS, EDF, and DREAM, respectively, with only $2.24\%$ average normalized accuracy loss on models with variants.

\bibliographystyle{IEEEtran}
\bibliography{references/reference}

\end{document}